\documentclass{mem}

\usepackage{natbib}
\usepackage{txfonts}
\usepackage{balance}
\usepackage{graphicx}
\usepackage[a4paper]{hyperref}
\idline{00}{159}

\begin{document}

\title{NGC 4245: one or two bars, and where does the 
  gas inflow stop?}

\subtitle{}

\author{O. K. Sil'chenko\inst{1} \and 
  I. V. Chilingarian\inst{1,2} \and V. L. Afanasiev\inst{3}}
          
\offprints{Olga K. Sil'chenko; \mbox{\email{olga@sai.msu.su}}}
 
\institute{Sternberg Astronomical Institute, Moscow State University, 
 Moscow, Russia
\and
 Laboratoire d'Etude du Rayonnement et de la
 Mati\'ere (LERMA), Observatoire de Paris, Paris, France
\and
 Special Astrophysical Observatory, Russian Academy of Sciences,
 Nizhnij Arkhyz, Russia }

\authorrunning{Sil'chenko, Chilingarian, and Afanasiev}

\titlerunning{Bar effects in NGC 4245}

\abstract{We have studied stellar and gaseous kinematics as well as stellar population properties in the center of the early-type barred galaxy NGC~4245 by means of integral-field spectroscopy. We have found a chemically distinct compact core, more metal-rich by a factor of 2.5 than the bulge, and a ring of young stars with the radius of 300\,pc. Current star formation proceeds in this ring; its location corresponds to the inner Lindblad resonance of the large-scale bar. The mean age of stars in the chemically distinct core is significantly younger than the estimate by Sarzi et al. (2005) for the very center, within $R=0\farcs25$, made with the HST spectroscopy data. We conclude that the `chemically distinct core' is in fact an ancient ultra-compact star forming ring with radius less than 100\,pc which marks perhaps the past position of the inner Lindblad resonance.  In general, the pattern of star formation history in the center of this early-type gas-poor galaxy confirms the predictions of dynamical models for the secular evolution of a stellar-gaseous disk under the influence of a bar.
\keywords{galaxies: elliptical and lenticular, cD -- 
  galaxies: evolution -- galaxies: individual: NGC 4245} }

\maketitle{}

\section{Introduction}

Secular evolution of disk galaxies is a slow but persistent process
that changes the appearance and structure of a galaxy drastically on
timescales of a few Gyrs. The main drivers of secular evolution are
bars, and the final point where gas accumulates and star formation
proceeds is the center of a galaxy. So to study consequences of
secular evolution in disk galaxies, we must look at their centers; and
to choose the most promising objects where secular evolution has
certainly occured we must look at galaxy groups. The groups are good
for the secular evolution realisation because the mutual movements of
galaxies are not so fast as in clusters, and the space density of
galaxies is high, so tidal interaction provoking bar formation, and
matter redistribution in a disk, is provided.

\begin{figure*}[]
\resizebox{\hsize}{!}{\includegraphics{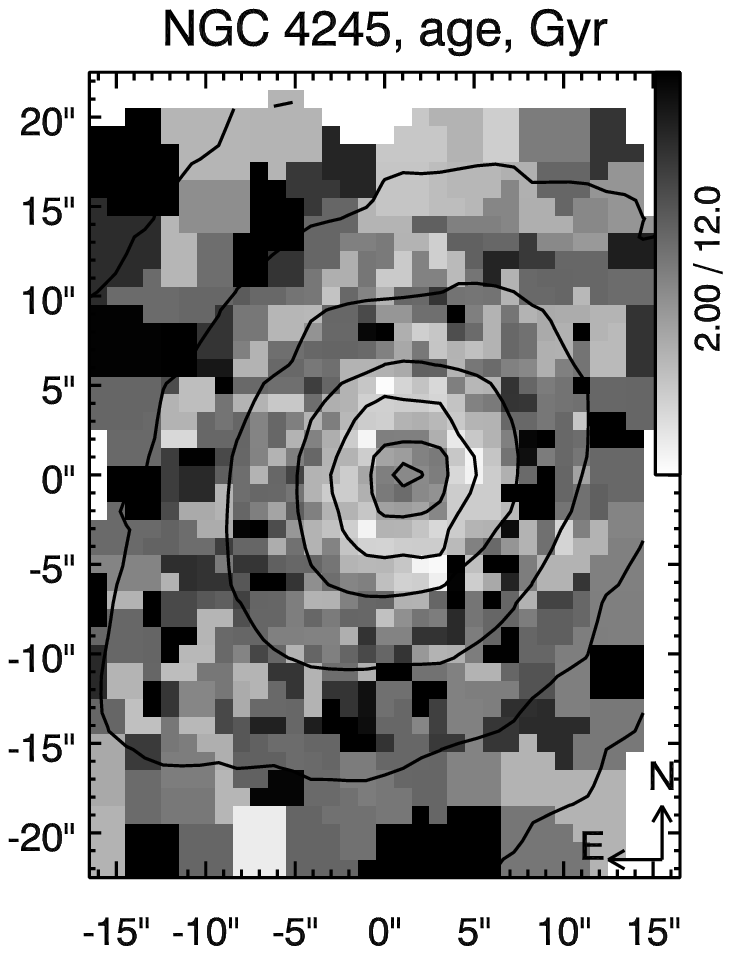}
\includegraphics{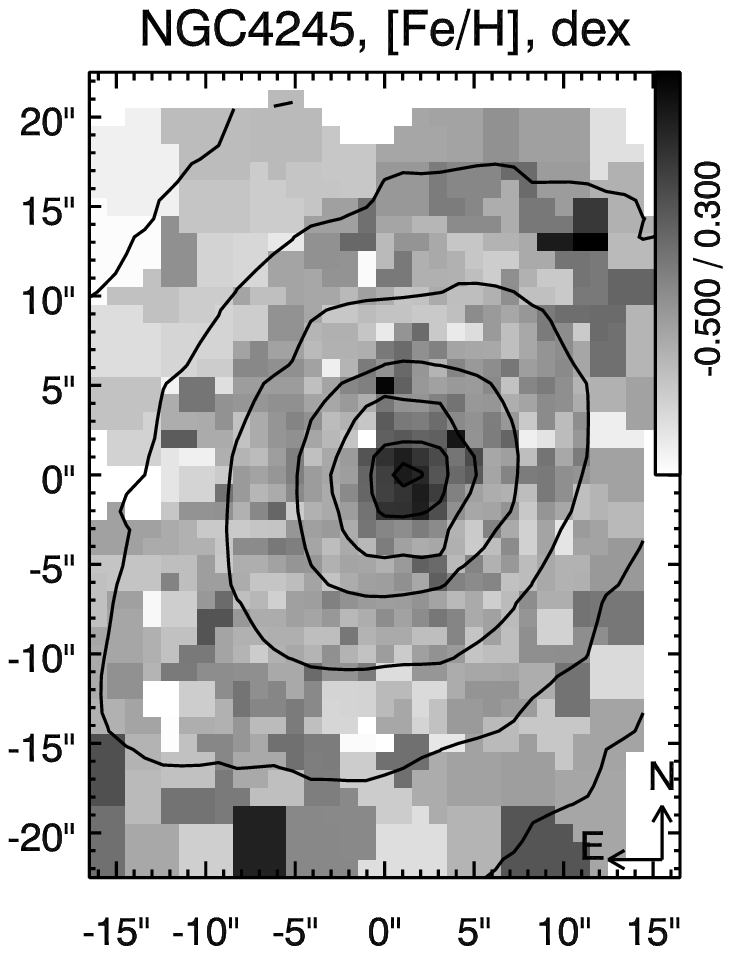}
\includegraphics{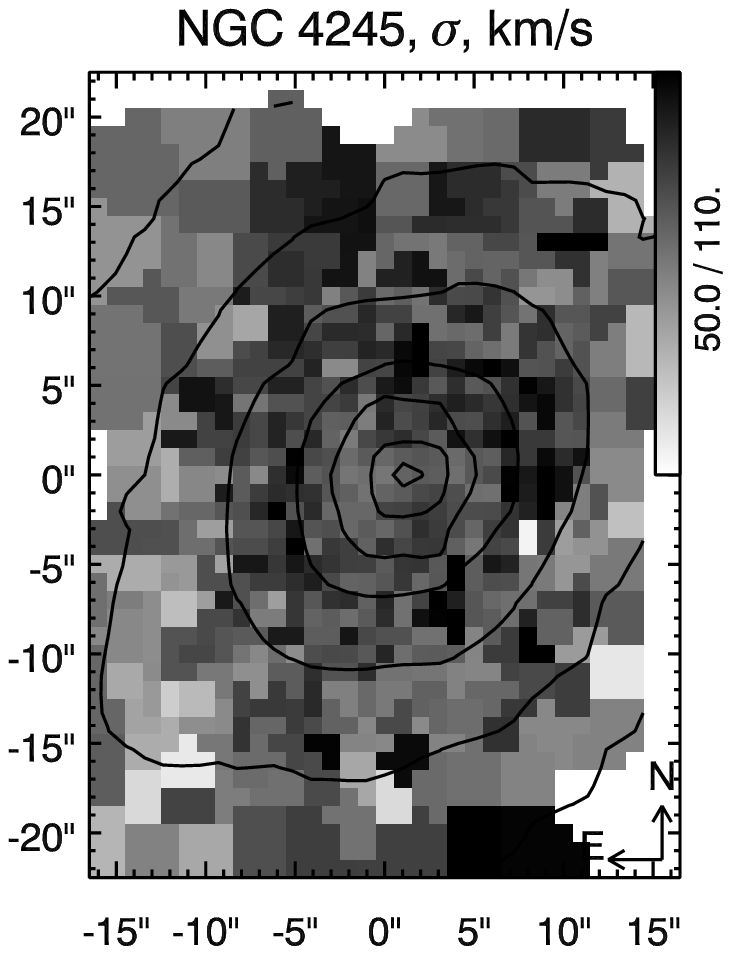}}
\caption{\footnotesize 
  Results of the fitting of the SAURON spectra for NGC~4245, from the
  left to the right -- SSP-equivalent stellar ages, SSP-equivalent
  stellar metallicity, and stellar velocity dispersion.  }
\label{maps}
\end{figure*}

NGC~4245 is an early-type disk galaxy (SB(r)0/a, according to
NED), of a moderate luminosity ($M_B=-18.74$ according to the
HYPERLEDA), with a large-scale strong bar -- a member of a rich,
spiral-dominated group Coma I Cloud \citep{grth}. NGC~4245 is located
in the dense part of the group, some 100\,kpc from the group center
and 59\,kpc from the giant spiral galaxy NGC~4274. As do all other disk
galaxies of the group Coma I Cloud, NGC~4245 demonstrates a strong
deficiency of neutral hydrogen \citep{comaihi}. The ratio of $M({\rm
H}_2) /M({\rm HI})\,\sim\,4$ in this galaxy \citep{comaimol}, and all
its gas is concentrated in the circumnuclear region where a
spectacular starforming ring is observed. Interestingly, the group
Coma I Cloud lacks X-ray intergalactic medium, so the common
explanation of the HI deficiency by the interaction with the hot
intergalactic medium is not valid in this case. It seems probable that
gravitational interactions play the main role in the group.  NGC~4245
as a recently formed lenticular galaxy may represent a late product of
secular evolution provoked by the strong bar and by the effects of a
dense environment.

\section{Observations}

We analyse here the integral-field spectral data obtained with the
Multi-Pupil Fiber Spectrograph (MPFS) of the Russian 6-m telescope and
the SAURON data retrieved from the ING Archive of the UK Astronomy
Data Center.  The former spectrograph \citep{mpfsman} gives a field
of view of $16^{\prime \prime} \times 16^{\prime \prime}$ and a
spectral range of 1500~\AA\ with the spectral resolution of 3~\AA;
the latter \citep{sau1} -- has a field of view of $41^{\prime \prime}
\times 33^{\prime \prime}$ with a spectral range of 550~\AA\ under
the spectral resolution of 4~\AA. The spectra have been fitted in the
pixel space by SSP models \citep{ic3653}, and the kinematic
characteristics as well as stellar population parameters have been
obtained.

\section{Results}

The maps of parameters of the stellar population in the central part
of NGC~4245 are presented in Fig.~\ref{maps}.  We can divide the
central part of NGC~4245 into three substructures differing by
kinematic and stellar population properties: the central unresolved
core which appears to be dynamically cold and chemically distinct, the
ring with the radius of $4^{\prime \prime} - 6^{\prime \prime}$ where
the emission line H$\alpha$ is extremely intensive and current star
formation proceeds, and the bulge without emission lines seen at
$R=7^{\prime \prime} - 10^{\prime \prime}$ where the maximum stellar
velocity dispersion is observed, the stars are old, and the
metallicity is subsolar. The comparison of the Lick indices measured
by us in these three substructures' MPFS spectra with the SSP models
of \citet{thomas} gives the following quantitative results: in the
core the metallicity is higher than solar by a factor of 3--5
and the mean stellar age is 2--4\,Gyr, in the ring the mean stellar age
is 1--2 Gyr, and in the bulge 8--12\,Gyr. The fitting of the SAURON
spectra with the PEGASE.HR code gives [m/H]$=+0.2$ and
$T_{\mbox{SSP}}=6.5$\,Gyr in the core, $T_{\mbox{SSP}}=2-4$\,Gyr in the
ring, and [m/H]$=-0.2$ and $T_{\mbox{SSP}}=8$\,Gyr in the bulge.

\section{Discussion}

\begin{figure}[]
\resizebox{\hsize}{!}{\includegraphics[clip=true]{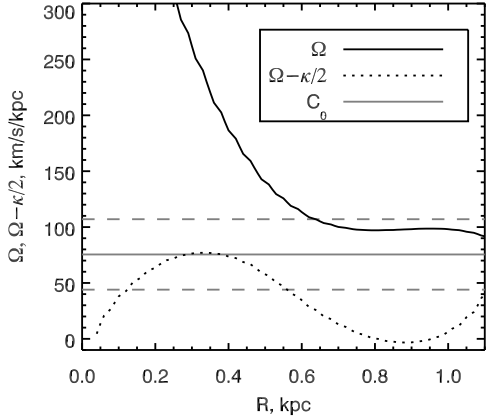}}
\caption{\footnotesize
  Positions of the inner Lindblad resonances derived from the analysis
  of the SAURON line-of-sight velocity map for NGC~4245.  }
\label{ilr}
\end{figure}

Starforming rings are commonly related to inner Lindblad resonances
(ILR) of bars \citep{bcreview}. The observational statistics of ring
sizes by \citet{bc93} has established this relation
empirically. Later simulations confirmed that gas inflowing along a
bar must accumulate just at ILRs \citep{hs96}. The starforming ring in
NGC~4245 answers all expectations.  Fig.~\ref{ilr} helps to determine
the ILR radii in this galaxy. We have taken the pattern speed of the
bar in NGC~4245 from \citet{trw4245}. With this pattern speed, the
galaxy may have two ILRs: the inner one at $R=5^{\prime \prime}$ and
the outer one at $R=6^{\prime \prime}$. The starforming ring is
observed just between them as theory predicts
\citep{piner,bcreview}.

\begin{figure}[]
\resizebox{\hsize}{!}{\includegraphics[clip=true]{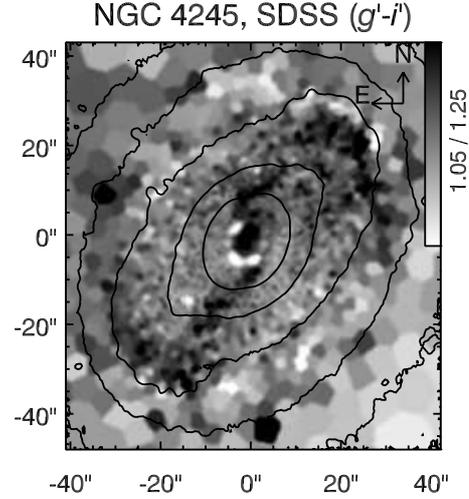}}
\caption{\footnotesize
  $g'-i'$ colour map of NGC~4245 built from the SDSS direct images.  }
\label{sdsscolour}
\end{figure}

Fig.~\ref{sdsscolour} demonstrates the $g^{\prime} -i^{\prime}$ colour
map of NGC~4245 constructed by using the SDSS data. Here we see also a
classical signature of gas response to the bar perturbation: thin
straight red (dust) lanes border the bar along its whole extension
betraying the presence of shocks produced by orbit crowding in the
triaxial potential \citep{atha}. Dust fronts end at the starforming
ring which is blue at the colour map. The blue starforming ring has
three knots which are the bluest and coincide with the H$\alpha$
condensations -- the sites of especially intense star formation; two
of them are just at points where the shock fronts meet the ring. Such
a picture is typical for star forming rings within bars -- see for
example a sample by \citet{rings2d}. Theoreticians explain this
phenomenon as follows: external disk gas inflows along the bar, meets
the dense gas concentration in the ring, produces shocks and provokes
star formation at the points of contact, and then these star forming
sites are driven by rotation into a complete ring.

According to theory, the gas inflow must stop at ILRs, and no gas
supply is allowed inside the ILRs. However our results give evidences
that some time ago gas inflow has overcome this obstacle.  The stellar
core is unresolved with MPFS and SAURON, and so the stellar
population within $1^{\prime \prime}-2^{\prime \prime}$ from the
center, is chemically distinct and is also significantly younger than
the bulge. So we conclude that there has been secondary star formation
burst {\em inside} $R=1^{\prime \prime}-2^{\prime \prime}$, or inside
the {\em present} ILRs, and this star formation burst has needed gas
supply to occur. But curiously, this star formation burst was also
annular, not nuclear: \citet{sarzi}, by analysing the HST spectral
data obtained within the aperture of $0\farcs25$,
determined for the nucleus of NGC~4245 the stellar age of 12\,Gyr and
no signs of gas emission. So the ancient star formation burst took
place in the ring between $0\farcs25$ ($R=15$\,pc) and
$1\farcs5$ ($R=100$\,pc).  Did this ring mark the previous
location of the ILR? In such case the chemically distinct core found by us
maybe a remnant of the starforming ring at the ILR of the (now)
dissolved inner bar and may be classified as an `ultracompact nuclear
ring' similar to those which have been discovered recently in some
barred galaxies \citep{ucsfrings}. Or may be the nuclear star-forming
ring in NGC~4245 changes its radial position as some theoretical
considerations predict \citep{regteub}? Then, since we know
the ages of the SF-rings at $R=300$\,pc and at $R=20-100$\,pc,
it would be interesting to try to relate possible interactions with
neighbouring galaxies with the ring shifts along the radius.

\begin{acknowledgements}
The work is supported by the grant of the Russian Foundation for Basic
Researches no. 07-02-00229a.  During the data analysis we have used
the Lyon-Meudon Extragalactic Database (HYPERLEDA) supplied by the
LEDA team at the CRAL-Observatoire de Lyon (France) and the NASA/IPAC
Extragalactic Database (NED) which is operated by the Jet Propulsion
Laboratory, California Institute of Technology, under contract with
the National Aeronautics and Space Administration.  This research is
partly based on data obtained from the Isaac Newton Group Archive
which is maintained as part of the CASU Astronomical Data Centre at
the Institute of Astronomy, Cambridge, and on SDSS data.  Funding for
the Sloan Digital Sky Survey (SDSS) and SDSS-II has been provided by
the Alfred P. Sloan Foundation, the Participating Institutions, the
National Science Foundation, the U.S. Department of Energy, the
National Aeronautics and Space Administration, the Japanese
Monbukagakusho, and the Max Planck Society, and the Higher Education
Funding Council for England. The SDSS Web site is
http://www.sdss.org/.
\end{acknowledgements}


\bibliographystyle{aa}

\begin{thebibliography}{}

\bibitem[Afanasiev et al.(2001)]{mpfsman}
  Afanasiev V.\ L., Dodonov S.\ N., \& Moiseev A.\ V.  2001, in
  Stellar dynamics: from classic to modern, ed. L. P. Osipkov, \&
  I. I. Nikiforov, (Sobolev Astronomical Institute, Saint Petersburg),
  103

\bibitem[Athanassoula(1992)]{atha}
 Athanassoula, E.\ 1992,  MNRAS,  259, 345

\bibitem[Bacon et al.(2001)]{sau1}
  Bacon, R., Copin, Y., Monnet, G., et al.\ 2001, MNRAS, 326, 23

\bibitem[Boker et al.(2008)]{rings2d}
  B\"oker, T., Falcon-Barroso, J., Schinnerer, E., Knapen, J.~H., \&
  Ryder, S.\ 2008, \aj., 135, 479

\bibitem[Buta \& Crocker(1993)]{bc93}
  Buta, R., \& Crocker D.~A.\ 1993, \aj, 105, 1344

\bibitem[Buta \& Combes(1996)]{bcreview}
  Buta, R., \& Combes, F.\ 1996, Fundam. Cosmic Phys., 17, 95

\bibitem[Chilingarian et al.(2007)]{ic3653}
  Chilingarian, I.~V., Prugniel, P., Sil'chenko, O.~K., \& Afanasiev,
  V.~L.\ 2007, MNRAS, 376, 1033

\bibitem[Comeron et al.(2008)]{ucsfrings}
  Comer\'on, S., Knapen, J.~H., Beckman, J.~E., \& Shlosman, I.\ 2008,
  \aap, 478, 403

\bibitem[Garcia-Barreto et al.(1994)]{comaihi}
  Garcia-Barreto, J.~A., Downes, D., \& Huchtmeier, W.~K.\ 1994, \aap,
  288, 705

\bibitem[Gerin \& Casoli(1994)]{comaimol}
  Gerin, M., \& Casoli, F.\   1994, \aap,  290, 49

\bibitem[Gregory \& Thompson(1977)]{grth}
  Gregory, S.~A., \& Thompson, L.~A.\ 1977, \apj, 213, 345

\bibitem[Heller \& Shlosman(1996)]{hs96}
  Heller, C.~H., \& Shlosman, I.\  1996, \apj,  471, 143

\bibitem[Piner et al.(1995)]{piner}
  Piner, B.~G., Stone, J.~M., \& Teuben, P.~J.\ 1995, \apj, 449, 508

\bibitem[Regan \& Teuben(2003)]{regteub}
  Regan, M.~W., \& Teuben, P.\ 2003, \apj, 582, 723

\bibitem[Sarzi et al.(2005)]{sarzi}
  Sarzi, M., Rix, H.-W., Shields, J.~C., et al.\ 2005, \apj, 628, 169

\bibitem[Thomas et al.(2003)]{thomas}
  Thomas, D., Maraston, C., \& Bender, R.\ 2003, MNRAS, 339, 897

\bibitem[Treuthardt et al.(2007)]{trw4245}
  Treuthardt, P., Buta, R., Salo, H., \& Laurikainen, E.\ 2007, \aj,
  134, 1195


\end{thebibliography}

\end{document}